\documentclass[%
 reprint,
 amsmath,amssymb,
 aps,
 prx,
]{revtex4-2}
\usepackage{physics} 
\usepackage{graphicx}
\usepackage{dcolumn}
\usepackage{bm}

\begin{document}

\preprint{APS/123-QED}

\title{ODMR on Single TR12 Centers in Diamond}

\author{Jonas Foglszinger$^1$}
\author{Andrej Denisenko$^1$}%
\author{Thomas Kornher$^2$}%
\author{Matthias Schreck$^3$}%
\author{Wolfgang Knolle$^4$}%
\author{Boris Yavkin$^5$}%
\author{Roman Kolesov$^1$}%
\author{Jörg Wrachtrup$^1$}%
\affiliation{$^1$3rd Institute of Physics, University of Stuttgart,70569 Stuttgart, Germany}%
\affiliation{$^2$Lumiphase AG, Brinerstrasse 21, 8003 Zürich, Switzerland}%
\affiliation{$^3$University of Augsburg, Institute of Physics, D-86135 Augsburg, Germany}
\affiliation{$^4$Leibniz Institute for Surface Engineering (IOM), Department Functional Surfaces, D-04318 Leipzig, Germany}
\affiliation{$^5$Quantronics Group, SPEC, CEA, CNRS, Université Paris-Saclay, 91191 Gif-sur-Yvette CEDEX, France}%
\date{\today}
             
\begin{abstract}
Point defects in insulators are considered promising candidates for quantum technologies.
In keeping with this, we present an extensive optically-detected magnetic resonance (ODMR) study at room-temperature on individual TR12 centers (ZPL at 471\,nm), which are known in the literature since 1956.
In this work we found TR12 centers to show a strong ODMR signal under optical saturation. 
These observed defects were created in high-purity epitaxial layers of diamond by standard irradiation and annealing processes.
From the analysis of the ODMR spectra along with antibunching measurements and coherent population trapping, we proposed the energy level structure of TR12 center, consisting of ground state and excited state singlets complemented by a metastable triplet in between.
Mapping the fluorescence dependence of the center on an external magnetic field and on the polarization of laser excitation, allows us to identify twelve inequivalent orientations for TR12 centers. 
This includes the exact orientations of the dipole transition and the triplet axes in the diamond lattice  in full agreement with the results of modeling based on the proposed level structure.
Furthermore, a static Jahn-Teller effect was detected through fluorescence switching between two levels at low optical excitation power, directly observable in the real-time fluorescence signal for various polarization of laser excitation.
Based on these results we discuss the prospects of the TR12 center in diamond for quantum sensing and quantum information processing.
\end{abstract}

\maketitle

\section{\label{sec:Intro}Introduction}
Over the past decades, the field of quantum technologies has become a rapidly developing area of research.
Solid-state host systems such as diamond and silicon carbide are particularly attractive in this respect as they inherit a rich variety of well-developed techniques from the electronic industry, such as material growth and processing,
implantation and photo-lithography. 
These solid-states offer access to different promising color centers such as nitrogen-vacancy (NV) centers in diamond \cite{NVC1,NVC2,NVC3} and silicon-vacancy (SiV) centers in Diamond \cite{SiV1, SiV2} or silicon carbide \cite{SiV3,SiV4}. 
As a pioneer, the NV center in diamond has already demonstrated its applicability in quantum computing and quantum sensing \cite{NV_sense_mag,SNV_sense_el,SNV_sense_temp,SNV_sense_Qcomp}, motivating further research with other color centers to explore different application areas, e.g. SiV, GeV\cite{GeV} or SnV\cite{SnV}.
Even though there are hundreds of color centers known in diamond \cite{ZaitsevsBook}, most of them do not facilitate spin control. 
Another pre-requisite for the center to be useful is that it can be created artificially at the location of interest. 
Here, we present the first ODMR studies on TR12 center exhibiting strong zero-phonon emission line (ZPL) at 471\,nm. 

The TR12 center is a point defect in diamond discovered in 1956 after electron or neutron bombardment \cite{TR12_discovery}. 
Its properties were investigated during the late 20th century in bulk measurements \cite{TR12_untersuch_1977, TR12_untersuch_1981, Mainwood_1994}. 
An existence as high density ensembles combined with a negative correlation involving nitrogen suggests TR12 to be an intrinsic defect. 
The existence of interstitial atoms is also supported by the appearance of a local phonon mode \cite{Mainwood_1994}.
However, the exact structure of the defect is still unknown.
Our studies were performed on single TR12 defects under the condition of optical saturation and allowed to reveal twelve inequivalent orientations of the center along with coherent properties of its associated spin in the metastable triplet state. 
We also show that TR12 defect exhibits a static Jahn-Teller distortion leading to switching between two spatial configurations manifesting themselves as slight changes in the orientation of the local defect symmetry.
at high optical excitation power.
\section{\label{sec:Experim}Experimental results}
\begin{figure*}
     \centering
     \includegraphics[width=\textwidth]{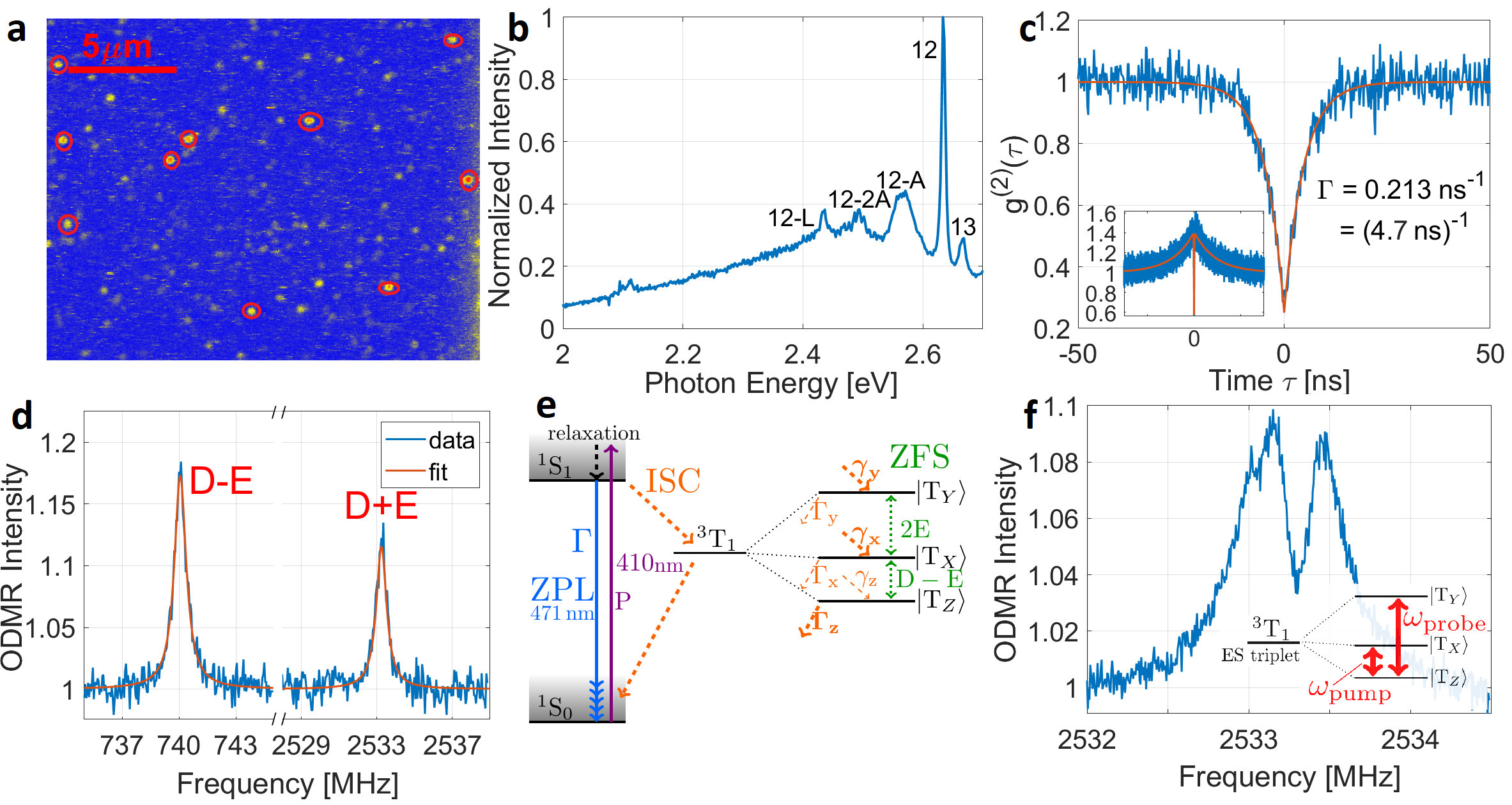}
        \caption{(a) Confocal scan image (20\,$\mu$m x 20\,$\mu$m) of the sample. Most of the optically active defects visible is the image are TR12 defects, some of which are marked by red circles. 
        (b) Fluorescence spectrum taken from a single TR12 center at room temperature, revealing a sharp ZPL along with several distinct phonon side bands. 
        (c) Antibunching measurement well below saturation on a single TR12 center with dip at $\tau=0$ below 0.5. 
        The inset shows a corresponding long-lasting  measurement at optical saturation revealing pronounced 'bunching-shoulder'. 
        (d) ODMR spectrum in zero field. The contrast can reach up to 30\% at high MW power for both observed lines $D-E$ and $D+E$. 
        (e) Proposed level-structure for the TR12 defect with ground state and excited state singlet complemented by a metastable triplet in between. 
        (f) CPT resonance indicating the existence of two long-lived states $T_x$ and $T_y$ \cite{Suppl}.}
        \label{fig:1,all}
\end{figure*}
\subsection{Sample preparation and experimental setup}
TR12 centers were created by either 10/370\,keV $^{12}$C ion implantation \cite{Roman_TR12} or 10\,MeV electron irradiation into (100) plane of CVD diamond followed by annealing at 800$^\circ$C for 1h \cite{TR12_tmepdep}.
While using $^{12}$C ions allows for exact positioning, these shallow implanted centers mostly lack the necessary photo-stability. 
Using higher implantation energies increases the average lifetime, but does not completely solve the problem.
Centers created by electron irradiation have higher photo-stability but are randomly distributed over the sample.
Beside the photo-stability, no spectroscopic difference was found between centers created by different methods.
The presented data is therefore acquired from defects within both, $^{12}$C ion implanted and electron irradiated samples.
Created defects were spectroscopically studied using a home-built confocal microscope with 410\,nm linear polarized laser excitation (\cite{Suppl} Fig. S1). 
All experiments were performed at room temperature.
The sample was scanned through the focal point by a 3D nano-positioner. 
The emitted fluorescence was split into two paths and detected by two single photon detectors for
integral fluorescence and fluorescence auto-correlation measurements.
Alternatively, the fluorescence could be deflected into a spectrometer for identification and spectral characterization of the emitters. 
Microwave (MW) radiation was supplied to the sample by a golden MW waveguide lithographically defined on the surface of the diamond. 
The setup was also equipped with a permanent magnet having magnetization perpendicular to the sample surface. 
Its position above the sample is controlled by high precision stepper motors. 
This allows the application of a magnetic field of variable magnitude and direction.

\subsection{ODMR of TR12 centers in zero field}
\begin{figure*}
     \centering
	\includegraphics[width=\textwidth]{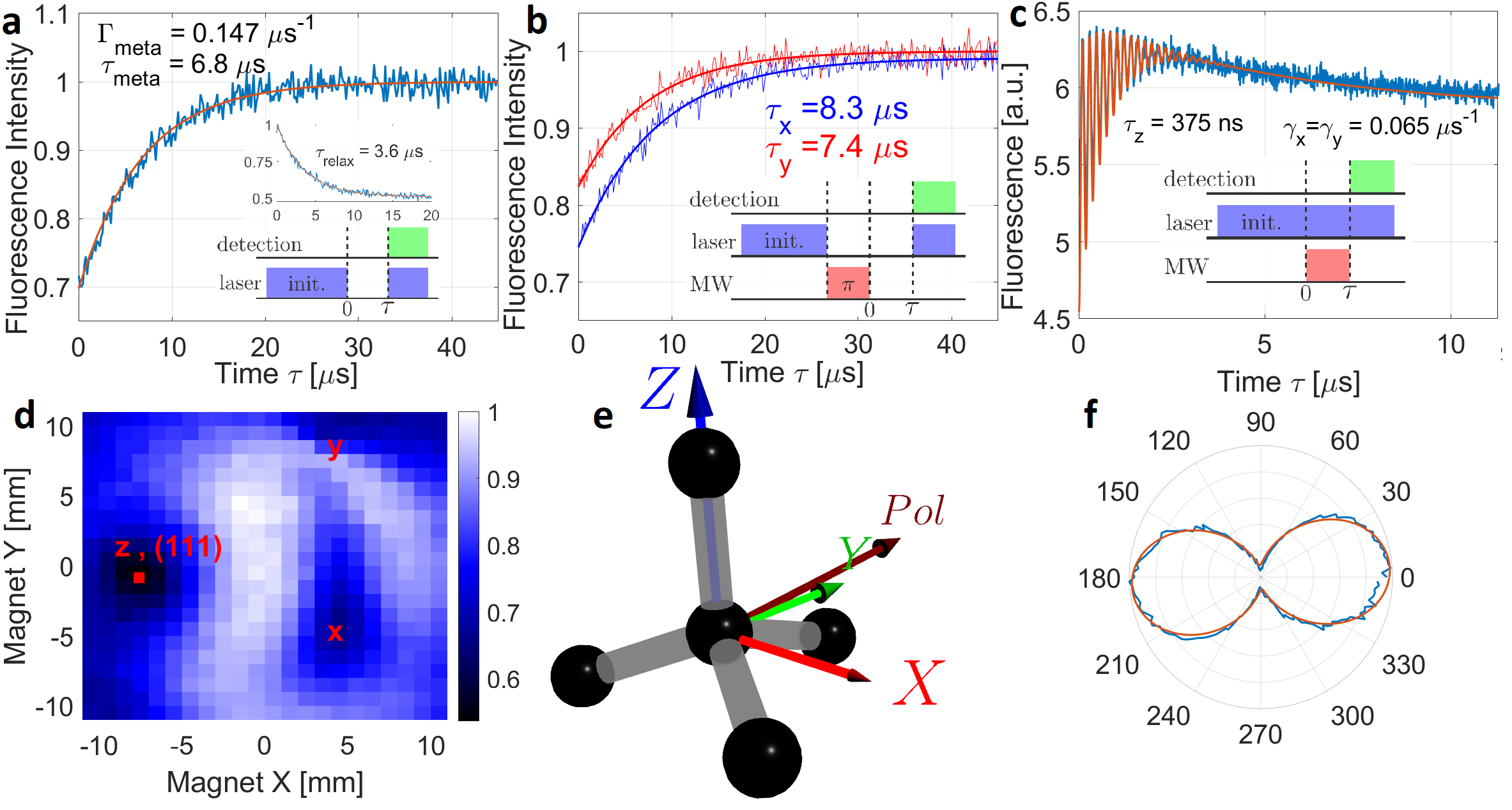}
             \label{fig:2,full}
             \caption{(a) Decay of the total metastable population. 
             The inset shows the time-dependence of fluorescence intensity, when the metastable states are repopulated after initialization in the ground state. 
             (b) Decay of metastable states $T_x$ and $T_y$. 
             (c) Rabi oscillation with full model fit \cite{Suppl}. 
             (d) Measured magnetic map for a TR12 center with marked (111) orientation in diamond 
                 and orientations for metastable triplet. 
             (e) Illustration of TR12 triplet orientations and the fluorescing transition dipole vector in the diamond lattice. 
             (f) Polarization dependent fluorescence measurement for a single TR12 defect, revealing a single dipole transition.}
\end{figure*}
TR12 defects were identified in a confocal scan (see Fig. 1a) by measuring their emission spectra (see Fig. 1b). 
Their single nature was confirmed by the fluorescence anti-bunching signal shown in Fig. 1c. 
The auto-correlation signal at low excitation power yields an estimate of the emitting state lifetime of 4.7\,ns. 
Under the condition of optical saturation of the defect, the fluorescence intensity strongly increases (up to a factor of 2) upon application of a magnetic field. 
Together with long-lasting photon bunching at saturating optical excitation (see inset of Fig. 1c), this
suggests the existence of a metastable optically populated electronic state.
As a result, ODMR could be observable for TR12 centers.
Indeed, in zero magnetic field two sharp ODMR lines can be observed at 740\,MHz and 2533.3\,MHz as shown in Fig. 1d. 
The ODMR contrast vanishes as the excitation laser power is reduced. 
To explain the observed ODMR signal, at least three distinct spin states must be considered. Based on this, the suggested electronic structure of the defect involves ground and excited state singlets responsible for the emission and a metastable triplet state in between (see Fig. 1e). 
The existence of exactly three distinct spin states in the metastable state was confirmed by  investigating the dependence of the fluorescence intensity on the magnetic field \cite{Suppl}.
The spin Hamiltonian describing the triplet state in zero magnetic field is given by 
$\textbf{H}=\mathrm{D}(S_z^2-1/3)+\mathrm{E}(S_x^2-S_y^2)$,
where $\mathrm{D}=1636.6\,\mathrm{MHz}$ and $\mathrm{E}=896.6\,\mathrm{MHz}$ can be deduced from the measured ODMR resonance frequencies. 
It results in three spin states $\ket{T_x}$, $\ket{T_y}$, and $\ket{T_z}$ and, therefore, in three possible ODMR transitions. 
In the experiment only two resonances were observed for different orientations for the center. 
Investigating centers on various positions around the MW antenna allowed to further manipulate the effective MW polarization without observing more than two resonances.
A wrong MW polarization can therefore be excluded as possible cause and two of the three spin states must share the same lifetime. 
Since the positive ODMR contrast is related to the redistribution of population from the long-lived occupied states to the short-lived empty states, there are two options: 
Either  $\ket{T_x}$ and $\ket{T_y}$ are occupied and have longer lifetimes than $\ket{T_z}$ or vice versa. 
By applying constant MW radiation at 740\,MHz and sweeping a second MW source through the second ODMR resonance, we observe coherent population trapping (CPT) as shown in Fig. 1f. 
This indicates that there are two long-lived states,  $\ket{T_x}$ and $\ket{T_y}$ \cite{Suppl}.
From the width of the CPT resonance the lifetime of these long-lived states can be estimated to be 5 to 10\,$\mu$s.
The lifetimes of the triplet sub-levels can also be measured directly by observing the decay of the shelving-state population. 
For this the laser excitation was interrupted for a variable time $\tau$ allowing the metastable population to partially decay to the ground state singlet. 
This results in a temporal increase of the fluorescence once the laser excitation is restored. 
The dependence of this temporal increase on $\tau$ reveals the metastable lifetime.
In Fig. 2a all states are monitored together, revealing only one exponential decay with a lifetime of 6.79\,$\mu$s fitting the estimation for the long-lived states $T_x$ and $T_y$. 
Consequently, the rate of intersystem crossing (ISC) into $\ket{T_z}$ has to fulfill  $\gamma_z\approx 0$, as $T_z$ must be negligibly populated.
To measure the lifetimes of $T_x$ and $T_y$ individually, one of them must be depleted by transferring its population to the fast decaying state $T_z$ with a MW $\pi$-pulse (see Fig.2b). 
In turn, the duration of a $\pi$-pulse is determined from MW-induced Rabi oscillations (see Fig.2c).
The resulting lifetimes roughly fit the expected value from the overall decay. 
Fig. 2b does only show a single exponential decay although the short-lived state $T_z$ must have been populated. 
\begin{figure*}
     \centering
         \includegraphics[width=\textwidth]{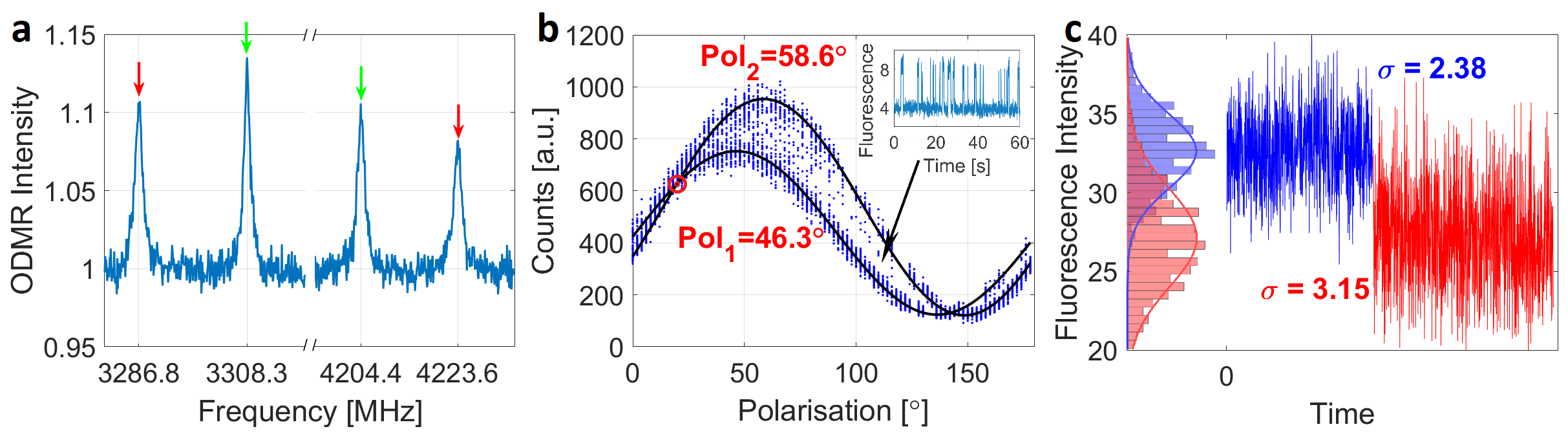}
     \caption{(a)  Anomalous splitting for both ODMR resonances as a result of the centers local symmetry axes turning with respect to the magnetic field. 
     Assigning the lines as indicated leads to the same magnetic field strength of 127.58\,mT ($\pm$0.01\,mT).
     (b) Polarization dependent fluorescence, supplemented by two fitted sine curves. 
     The inset shows a lifetime trace from a single TR12 center for polarization of high contrast switching as indicated. 
     The red circle marks the polarization angle of equal brightness at which the adjacent Fig. 3c was taken.
     (c) Fluorescence trace with two active MW sources (blue) and one active MW source (red).
      The overall noise increases although the shot noise is reduced with the fluorescence amount.
      This increased noise is due to the switching between two different fluorescence levels which can not be resolved in time at high optical excitation power.}
\end{figure*}
This is caused by an additional offset in time which was added deliberately to cut off the low resolution data from this decay. 
Instead $\tau_z$ was determined indirectly, by fitting it to the decay of the Rabi-oscillation between $T_x$ and $T_z$. 
To account for the overall dynamics of the system, a full model simulation \cite{Suppl} has been fitted to the data in Fig. 2c. 
By making the additional assumption about ISC rates into $T_x$ and $T_y$  $\gamma_x=\gamma_y$, not only the lifetime $\tau_z$, but also these population rates can be extracted. 
This can be understood by considering the different timescales on which the respective effects take place. 
Rabi-oscillations decay within about $1\,\mu$s, from where $\tau_z=375\,$ns is fitted. 
On the other hand Fig. 2c shows that on a timescale of 5 to 10\,$\mu$s, a new steady fluorescence level is set. 
As this level is mostly defined by the rates into the metastable state $\gamma_x=\gamma_y=0.065\,\mu \mathrm{s}^{-1}$, these values can be estimated here. 
Simulations considering the overall ODMR contrast, lead to the more precise values $\gamma_x=\gamma_y=0.095\,\mu \mathrm{s}^{-1}$.

\section{\label{sec:Mag} Fluorescence dependence on the magnetic field}
Studying the dependence of the fluorescence on the external magnetic field provides further confirmation of the electronic level structure and gives insight into the symmetry of TR12 defects. 
For that, a permanent magnet (10x10x10\,mm$^3$ NdFeB, magnetization 1.4\,T perpendicular to the sample surface) is moved atop of the sample while the fluorescence intensity of the center is recorded depending on the magnet position.
The dependence of the fluorescence on the magnetic field comes from mixing of the least metastable $T_z$ to $T_x$ and/or $T_y$ modifying their relaxation rates.
This mixing is described by the Hamiltonian
$\textbf{H}=\textbf{S}*\textbf{D}*\textbf{S}+g\mu_B \textbf{S}*\textbf{B}$
where $g$ is the g-factor and $\mu_B$ the Bohr magneton.
An exemplary magnetic map is given in Fig. 2d.
The dark spots on the map correspond to the orientations of the magnetic field not leading to mixing of $T_z$ with at least one of the $T_x$ and $T_y$ states. 
The darkest spot marks the magnetic field pointing along the local $z$-axis of the center while the other two reveal the orientation of $x$- and $y$-axes. 
The results were compared to magnetic maps taken for NV centers in the same sample and revealed the $z$-axis being oriented close to (111) orientation in diamond \cite{Suppl}. 
Simulations of the magnetic maps showed that the $y$-axis of the TR12 is lying in the plane made by two adjacent $\sigma$-bond. 
This fully defines the local frame of the center since the $x$-axis is perpendicular to $z$ and $y$ (see Fig. 2e). 
From the symmetries within the diamond lattice combined with simulations, we propose twelve inequivalent orientations for TR12 which were also confirmed in measurements.
The corresponding magnetic maps are listed in the Supplementary Information \cite{Suppl}.
Lastly, we studied the dependence of the fluorescence on the polarization of excitation light. 
TR12 center has only one optical dipole as can be seen from the polarization-dependent fluorescence pattern shown in Fig. 2f. 
Even though only the projection of the dipole on the plane of the sample is measured directly, combining the projections for twelve magnetically inequivalent species allowed for unambiguous determination of the dipole direction in the local frame of the TR12 center. 
It appears to be slightly tilted with respect to the $y$-axis (see Fig. 2e).
\section{\label{sec:JT} Jahn-Teller effect}
An evidence for strong coupling between electronic and vibrational degrees was presented in an earlier work by Davies \cite{TR12_untersuch_1981}. 
In the current work, we are providing further evidence to this complexity by demonstrating the following effects, which could eventually lead to identifying the structure of the TR12 center. 
First, we observe a splitting of the ODMR lines of  single TR12 into two under any external magnetic field which is identical for all centers sharing the same orientation.
This cannot be explained by  coupling to a nuclear spin since the number of split lines stays two at arbitrary orientation of the magnetic field. 
Indeed, nuclear spin-flip transitions should be allowed for arbitrarily oriented field and, therefore, for nuclear spin $\mathrm{I}=1/2$ one expects four resonances while we observe only two. 
At zero magnetic field no splitting could be observed. 
The second observation was made at low optical excitation, when telegraphic switching of the fluorescence intensity between two well defined levels was observed (see Fig. 3b inset). 
The dependence of the telegraphic signal on the polarization of laser excitation is shown in Fig. 3b. 
One can see two sinusoidal signals shifted by about $12^{\circ}$ with the fluorescence intensity switching between them.

Both phenomena occurring jointly together can be explained if the TR12 center is hopping between two stable energetically degenerate configurations whose local symmetry axes are tilted with respect to each other. 
On the one hand, the optical dipole projection turns by about twelve degrees leading to different efficiency of laser excitation for a given polarization, which explains the telegraphic fluorescence signal. 
The switching becomes faster as the excitation power increases and is barely resolvable once the center is saturated. 
On the other hand, splitting of the ODMR line is caused by turning of the symmetry axes of the center with respect to the magnetic field. 
If both split ODMR lines are measured, one can always assign two pairs of resonances corresponding to exactly the same magnetic field strength (127.58\,mT in Fig. 3a) meaning that the center is turning with respect to the B-field.

As the fluorescence switching can only be resolved at low excitation power, while ODMR does only appear upon fluorescence saturation, there is no way to synchronize ODMR and switching.
A way to nevertheless link these two effects, is to set the polarization of the laser such that both dipoles make the same angle with the laser polarization. 
In this case, no switching is observable with none of the two split transitions or both driven by MWs since both configurations are equally bright as indicated in Fig. 3b. 
However, if MWs are applied in resonance with only one of the two split transitions, switching is restored as only one of the configurations contributes to the ODMR signal. 
Since switching is fast under optical saturation, this effect can be seen only as increased noise (see Fig. 3c).

\section{\label{sec:CO} Conclusion and outlook}

\begin{figure}
     \centering
         \includegraphics[width=0.45\textwidth]{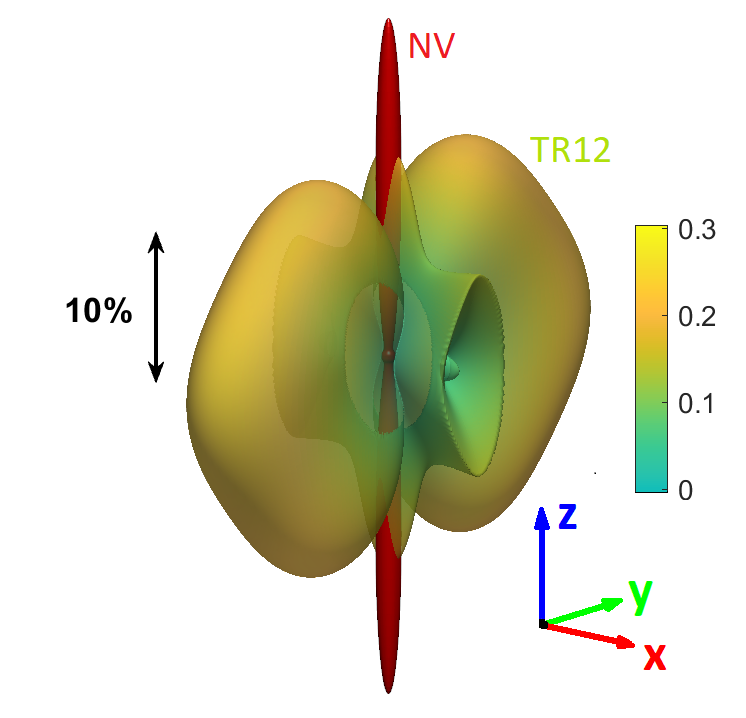}
     \caption{Observable ODMR contrast as a function of the magnetic field orientation for TR12 (multicolored) and NV (red) for 30\,mT. 
     Since two resonance frequencies are required to perform magnetic calculations, the plot displays the second highest contrast for every orientation. 
     Thus, for all orientations of the magnetic field, two resonance lines can be observed that show the contrast indicated in the plot and higher. The orientation of the TR12 metastable triplet is indicated by colored arrows.}
\end{figure}
The presence of spin-active state of the TR12 defect in diamond has been demonstrated by means of confocal ODMR spectroscopy. 
Detailed spectroscopic studies carried out on a large number of single defect centers allowed us to collect enough statistics to propose an adequate model for the defect to describe its intriguing optical and spin properties.
Twelve inequivalent orientations of the center in the diamond lattice are revealed by mapping its fluorescence as a function of external magnetic field. 
Additionally, we observed switching of the optical dipole of TR12 center between two distinct configurations and attributed this to static Jahn-Teller effect. 
This effect also represents itself in anomalous splitting of ODMR resonances in external magnetic field.

The TR12 defect can be an interesting alternative to the NV center in diamond in quantum sensing and quantum information processing. 
For sensing applications, TR12 center has its own pros and cons. 
It can be used to sense magnetic fields, temperature, strain, and, possibly, electric field, though its responsivity to temperature and electric field is still to be assessed.
On one hand, since ODMR is observed in the excited triplet state, the width of ODMR resonances is intrinsically limited by its lifetime. 
Therefore, the sensitivity is limited fundamentally. 
On the other hand, TR12 has a much larger acceptance angle for measuring high magnetic fields on the order of tens of milli-Tesla to Tesla. 
While NV center can be used to sense the fields in this range only along its symmetry axis, the TR12 shows non-zero ODMR contrast in a wide range of angles along its local $z$-axis and around its local $xy$-plane. 
Simulated ODMR contrasts as a function of magnetic field direction are presented in Fig. 4 and compared to the contrast of an NV center. 
Thus, TR12 can be used as a full vector nanoscale magnetometer for almost arbitrarily oriented B-field.

The defect also has some advantages over the NV center in quantum information processing. 
One can use it as a communication tool to initialize and read out nearby nuclear spin qubits (e.g. {$^{13}$C}) once in the triplet state. 
However, the nuclear spin memories will not be affected by its spin when TR12 is in the ground state singlet. 
Thus, a longer nuclear memory lifetime is expected \cite{NV_dec}. 
Furthermore, strong ZPL of TR12 if enhanced by Purcell effect in a microcavity can serve as an interface between flying qubits (photons) and stationary ones (nuclear spins) \cite{Purcell1}.
All this makes TR12 an interesting alternative and ally to the well-developed family of NV centers and IV-group divacancies for future studies and applications of solid-state optically active spin defects.

\section{Acknowledgments}
We thank Adam Gali for helpful discussions.
This work was supported by Bundesministerium für Bildung und Forschung (project UNIQ), Deutsche Forschungsgemeinschaft (grant KO4999/3-1), FET-Flagship Project SQUARE, ERC grant SMeL, EU project ASTERIQS, DFG research group FOR 2724 and QTBW.

\bibliography{Literatur}
\end{document}



\preprint{APS/123-QED}

\title{Supplemental Material on \\ODMR on Single TR12 Centers in Diamond}

\author{Jonas Foglszinger$^1$}
\author{Andrej Denisenko$^1$}%
\author{Thomas Kornher$^2$}%
\author{Matthias Schreck$^3$}%
\author{Wolfgang Knolle$^4$}%
\author{Boris Yavkin$^5$}%
\author{Roman Kolesov$^1$}%
\author{Jörg Wrachtrup$^1$}%
\affiliation{$^1$3rd Institute of Physics, University of Stuttgart,70569 Stuttgart, Germany}%
\affiliation{$^2$Lumiphase AG, Brinerstrasse 21, 8003 Zürich, Switzerland}%
\affiliation{$^3$University of Augsburg, Institute of Physics, D-86135 Augsburg, Germany}
\affiliation{$^4$Leibniz Institute for Surface Engineering (IOM), Department Functional Surfaces, D-04318 Leipzig, Germany}
\affiliation{$^5$Quantronics Group, SPEC, CEA, CNRS, Université Paris-Saclay, 91191 Gif-sur-Yvette CEDEX, France}%
\date{\today}

\maketitle
\section{Experimental setup}
\begin{figure}
     \centering
         \includegraphics[width=0.45\textwidth]{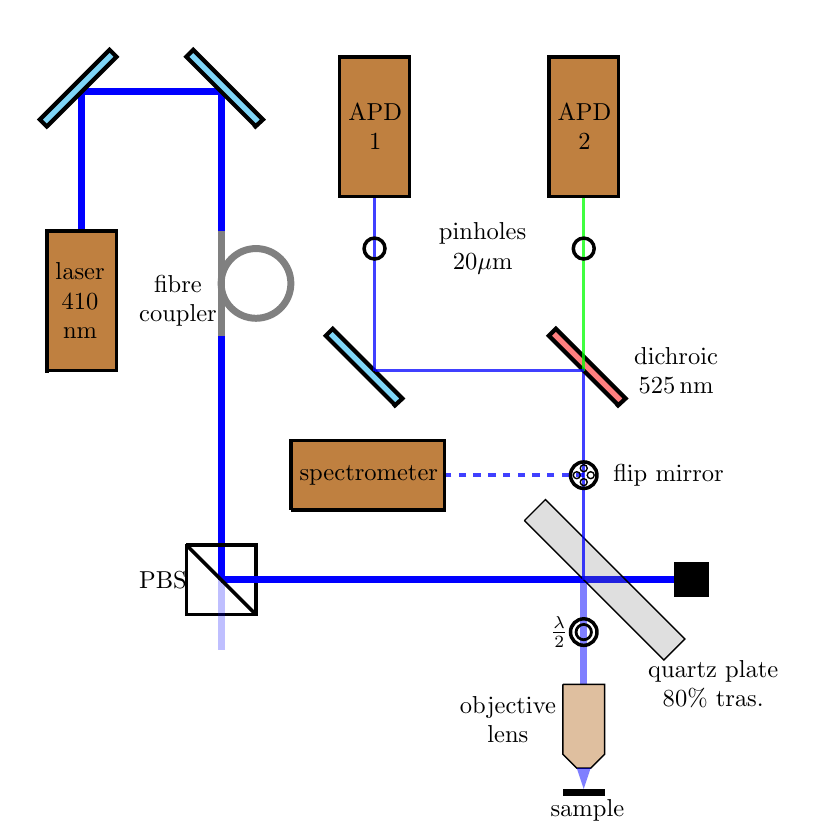}
     \caption{Schematic representations of the used home-built confocal microscope.}
\end{figure}
All experiments presented were performed using a home-build confocal microscope (see Fig. S1) working at room temperature. 410\,nm laser light is coupled into a single mode polarization
maintaining optical fiber to obtain Gaussian mode profile.
A polarizing beam splitter (PBS) ensures only one linear polarization is incident on the sample. 
Combined with the $\lambda$/2-plate right in front of the objective lens, this allows measurements dependent on excitation polarization.
The fluorescence light is collected via the same objective lens, split at a dichroic mirror and detected by two single photon detectors which are based on the principle of avalanche photo diodes (APDs).
Alternatively, the fluorescence can be deflected to a spectrometer. 
The collected light is sent through a 20\,$\mu$m pinhole to ensure high axial resolution.
\section{Coherent population trapping}
In this section we briefly introduce the basic theory of coherent population trapping (CPT) in a three-level system. 
The considerations only cover what is  needed for this paper. More extended analysis can be found in \cite{CPT2,CPT3,CPT4}, which is based on theoretical foundations such as \cite{CPT5,CPT6,CPT7}.

Consider a three-level system such as our exited triplet. 
The sates are labeled $\ket{T_x}$, $\ket{T_y}$ and $\ket{T_z}$ as in Fig. 1e. 
The corresponding Hamiltonian reads
\begin{equation}
\mathrm{H}_\mathrm{0}=\hbar\omega_x\ket{T_x}\bra{T_x}+\hbar\omega_y\ket{T_y}\bra{T_y}+\hbar\omega_z\ket{T_z}\bra{T_z}\,,
\end{equation}
were $\hbar\omega_i$ are the energies for the states $\ket{T_i}$.
Let $\ket{T_z}$ be the short-lived state while $\ket{T_y}$ and $\ket{T_z}$ are long-lived.
When both long-lived states are coupled to the short-lived via microwaves, this can be expressed as an additional Hamiltonian 
\begin{equation}
\mathrm{H}_\mathrm{1}=-\frac{\hbar}{2}\left(\Omega_{p}e^{-i\omega_{p}t}\ket{T_x}\bra{T_z}+\Omega_{c}e^{-i\omega_{c}t}\ket{T_y}\bra{T_z}\right) + \mathrm{H.c}
\end{equation}
where $\Omega_{p}$ and $\Omega_{c}$ are the two different Rabi frequencies for the transition frequencies $\omega_p=\omega_x-\omega_z$ and $\omega_c=\omega_y-\omega_z$.
Solving the Schrödinger equation for the full Hamiltonian $H = H_0+ H_1$ with the ansatz for the wave functions
\begin{equation}
\ket{\psi(t)}=c_x(t)e^{-i\omega_xt}\ket{T_x}+c_y(t)e^{-i\omega_yt}\ket{T_y}+c_z(t)e^{-i\omega_zt}\ket{T_z}\,,
\end{equation}
results in a simple set of differential equations for the coefficients $c_i$
\begin{equation}
\begin{split}
\dot{c_x}&=\frac{i}{2}\Omega_pc_z\\
\dot{c_y}&=\frac{i}{2}\Omega_cc_z\\
\dot{c_z}&=\frac{i}{2}\left(\Omega_pc_x+\Omega_cc_y\right)\,.\\
\end{split}
\end{equation}
We notice , that steady-state solution is: 
\begin{equation}
\begin{split}
c_x=\cos(\theta),\quad c_y=-\sin(\theta)\quad\mathrm{ and }\quad c_z=0\quad \mathrm{where}\\
\cos(\theta)=\frac{\Omega_c}{\sqrt{\Omega_c^2+\Omega_p^2}}\quad\mathrm{ and }\quad 
\sine(\theta)=\frac{\Omega_p}{\sqrt{\Omega_c^2+\Omega_p^2}}\,.
\end{split}
\end{equation}
In other words, there is a dark state $\ket{\psi_D}=\cos(\theta)\cdot\ket{T_x}-\sin(\theta)\cdot\ket{T_y}$, with no probability to be in $T_z$ or change to $T_z$ although $T_x$ and $T_y$ are both coupled to $T_z$ via microwaves.
The effect of population being trapped in this dark-state is called coherent population trapping.\\

The dark state, as superposition, inherits the long lifetime from $T_x$ and $T_y$. Consequently it has a rather narrow natural linewidth.
In the experiment shown in Fig. 1f, this results in a broad resonance peak with linewidth corresponding mostly to the lifetime of $T_z$ with an additional dip with linewidth corresponding to the long lived states $T_x$ and $T_y$.
To complete the argument from the main-text, one must realize, that this effect would not occur if the situation was reversed.
If the positive contrast in Fig 1.d would result from a long-lived state $T_z$ being coupled to one of the short-lived states $T_x$ or $T_y$, no such dark state could be formed, as any superposition making all derivatives to zero, would involve at least one short lived state. 
This superposition would therefore have short lifetime as well and would not allow any fluorescence quenching with narrow linewidth.
The existence of the narrow Fluorescence dip within the broader peak in Fig. 1f therefore indeed confirms the existence of two long lived states $T_x$ and $T_y$.
\section{Steady state solution}
The dynamics in the unperturbed system was simulated using the matrix equation
\begin{equation}
\dot{\rho}=M\rho
\label{steady_state}
\end{equation}
with the vector of states \\
$\rho=(\rho_g(\mathrm{S}_0), \rho_e(\mathrm{S}_1), \rho_x,\rho_y,\rho_z)^{tr}$ and a Matrix
\begin{equation}
M=\left(\begin{array}{c c c c c} 
-P&\Gamma&\Gamma_X&\Gamma_Y&\Gamma_Z\\
P& -\Gamma-\gamma_x-\gamma_y-\gamma_z&0&0&0\\
0&\gamma_x&-\Gamma_X&0&0\\
0&\gamma_y&0&-\Gamma_Y&0\\
0&\gamma_z&0&0&-\Gamma_Z\\\end{array}\right)
\label{steady_state_2}
\end{equation}
defined by the transition rates within the electronic level structure depicted in Fig. 1f. 
When combining this matrix equation with $\dot{\rho}=0$, the steady state solution can be calculated.
As the fluorescence is generated from the transition $\mathrm{S}_1\rightarrow{\mathrm{S}_0}$, the fluorescence intensity is directly proportional to the population of the excited state $\rho_e(\mathrm{S}_1)$.

\begin{figure*}
     \centering
         \includegraphics[width=\textwidth]{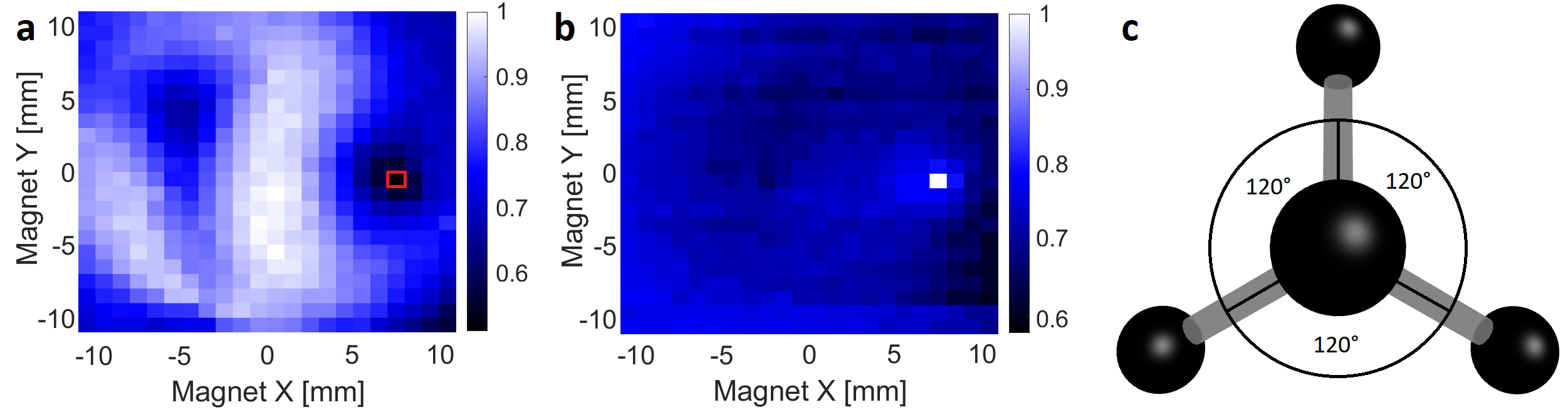}
     \caption{ (a) Measured magnetic map for TR12 center with marked NV $z$-orientation (111)
in diamond. (b)  Measured magnetic map for NV center used as a reference for orientation of the magnetic field.
     (c) Graphic display of the tetrahedron, to illustrate diamond lattice along (111) revealing a threefold symmetry. 
     After rotation by 120$^{\circ} $ the tetrahedron merges with itself.}
\end{figure*}
\section{Rabi oscillation}
In order to simulate the fluorescence during a Rabi oscillation, the equations in \ref{steady_state_2} were modified to include the respective off-diagonal elements needed.
Assuming the coherent driving to take place between the states $T_z$ and $T_x$, the result reads
\begin{equation}
\begin{split}
\dot{\rho_g}&=-P\cdot\rho_g+\Gamma\cdot\rho_e+\Gamma_x\cdot\rho_x+\Gamma_y\cdot\rho_y+\Gamma_z\cdot\rho_z\\
\dot{\rho_e}&=P\cdot\rho_g -(\Gamma+\gamma_x+\gamma_y+\gamma_z)\cdot \rho_e\\
\dot{\rho_x}&=\gamma_x\cdot\rho_e-\Gamma_x\cdot\rho_x + \frac{i\Omega}{2}(\sigma_{zx}-\sigma_{xz})\\
\dot{\rho_y}&=\gamma_y\cdot\rho_e-\Gamma_y\cdot\rho_y\\
\dot{\rho_z}&=\gamma_z\cdot\rho_e-\Gamma_z\cdot\rho_z + \frac{i\Omega}{2}(\sigma_{xz}-\sigma_{zx})\,\\
\dot{\sigma}_{xz}&=\frac{i\Omega}{2}(\rho_{z}-\rho_{x})-\sigma_{xz}\left(\frac{\Gamma_x}{2}+\frac{\Gamma_z}{2}\right)\\
\dot{\sigma}_{zx}&=\frac{i\Omega}{2}(\rho_{x}-\rho_{z})-\sigma_{zx}\left(\frac{\Gamma_x}{2}+\frac{\Gamma_z}{2}\right)\,.
\label{Rabi_sim}
\end{split}
\end{equation}
For the simulation in Fig. 2c, the system was first initialized in the steady state using equation \eqref{steady_state} and then evolved with the set of linear differential equations \eqref{Rabi_sim}. 
As the the fluorescence measurement does not take place instantaneously, the system further evolves during this measurement. 
This is taken into account in the simulation by also further evolving the system for the readout time without Rabi oscillation and averaging the brightness over this interval.

\section{NV center as reference for magnetic maps}
The problem of magnetic maps as given in Fig. 2d, is the lack of any reference. 
While it is obvious that changing the magnet position does change the magnetic field strength and orientation, the exact values can not be read out directly. 
Also simulating this orientation would not provide a proper solution, as the magnet was only centered above the sample rather coarsely which very much reduces the precision of this approach. 
Instead the process of taking magnetic maps was repeated for NV centers which are known to have their $z$-orientation aligned with (111) in diamond.
In a magnetic map for NV center, this orientation will appear as single bright spot serving as clear reference.
When marking the position of this spot in a magnetic map for TR12 center, a precise reference is given, which reveals the $z$-orientation of TR12 metastable triplet to be oriented along (111) in diamond as for NV (see Fig. S2a,b) with a maximum deviation of about 5$^\circ$.
\section{Simulating magnetic maps and orientations}
If the $z$-orientation of TR12 metastable triplet is known to be oriented along (111) in diamond, proposing the existence of twelve different orientations is rather straight forward.
In diamond lattice there are four distinct orientations for (111) which automatically transfers to four different $z$-orientations. 
Imaging the diamond lattice along (111) (See Fig. S2c) a threefold symmetry for rotations around $z$ by 120\,$^\circ$ becomes obvious.
Therefore there are three possible orientations for $x$ and $y$ sharing the same $z$-orientation. 
This results in twelve inequivalent orientations which are intuitively sorted as four triples.

In order to simulate magnetic maps, the magnetic field in the plane of interest below the magnet is simulated at first. 
The permanent magnet is approximated as a cube consisting of uniformly distributed magnetic dipoles.
This simulated magnet is then moved virtually above the sample as in the measurement and the resulting magnetic field calculated at the centers position.
As TR12 centers can be oriented differently, the orientation of the magnetic field in the local frame of the center has to be calculated.
To simulated the changes in brightness, the effect of this magnetic field on the metastable triplet state is determined.

When a static magnetic field is applied, the system Hamiltonian consists of Zero field splitting part and Zeeman interaction part
\begin{equation}
\textbf{H}=\textbf{SDS}+g_s\cdot\mu_B\cdot\textbf{SB},
\end{equation}
with g-factor $g=2$ and Bohr-magneton $\mu_B$.
The resulting matrix provides three eigenvectors $\varphi_i$ which depend on the magnetic field.
Expressing these eigenvectors as a linear combination of the eigenvectors resulting from the unperturbed Hamiltonian $\textbf{H}=\textbf{SDS}$ reads
\begin{equation}
\varphi_i=\alpha_i\cdot \ket{T_x}+\beta_i\cdot\ket{T_y}+\zeta_i\cdot\ket{T_z}\,.
\end{equation}
Assuming that all cross terms average to zero, the new transition rates into ($\gamma_i$) and out of ($\Gamma_i$) the metastable triplet can be calculated to be 
\begin{equation}
\begin{split}
\gamma_1=&\left|\bra{g}H_\mathrm{int}\ket{\varphi_1}\right|^2\\
=&\left|\bra{g}H_\mathrm{int}\ket{\alpha_1\cdot EV_x}\right|^2\\
&+\left|\bra{g}H_\mathrm{int}\ket{\beta_1 \cdot EV_y}\right|^2\\
&+\left|\bra{g}H_\mathrm{int}\ket{\zeta_1\cdot EV_z}\right|^2\\
=&|\alpha_1|^2\cdot\gamma_x+|\beta_1|^2\cdot\gamma_y+|\zeta_1|^2\cdot\gamma_z\,
\end{split}
\end{equation}
or analog to this. 
Using these new transition rates, the steady state solution can be calculated for every magnetic field orientation. 
As the reference given by the NV center only fixes the $z$-orientation of TR12 triplet, the orientation of $x$- and $y$-axis still had to be obtained. 
This was achieved by rotating $x$- and $y$-axis around $z$-orientation, always forming an orthogonal system.
The resulting magnetic map was then graphically fitted to the real measurement via this rotational angle.
With a clear match the $y$-orientation was found to lie in the plane formed by two sigma bonds also fixing $x$-orientation.
The respective local frames for the remaining eleven orientations were obtained by the described rotations by 120$^\circ$ and by reorienting the $z$-orientation along another sigma bond ((111) in Diamond).
Using these local frames lead to twelve simulated magnetic maps in Fig. S3. For comparison, all measured maps are listed in the same figure below.

From the fact that simulations and measurements match almost perfectly, it follows that the proposed electronic structure (Fig. 1f) is indeed an accurate description for TR12.
As the example of the magnetic map for an NV centre (Fig S2b) shows, a different electronic structure would lead to a completely different magnetic map.
This especially confirms the spin multiplicity of the metastable triplet.
\begin{figure*}
     \centering
         \includegraphics[width=\textwidth]{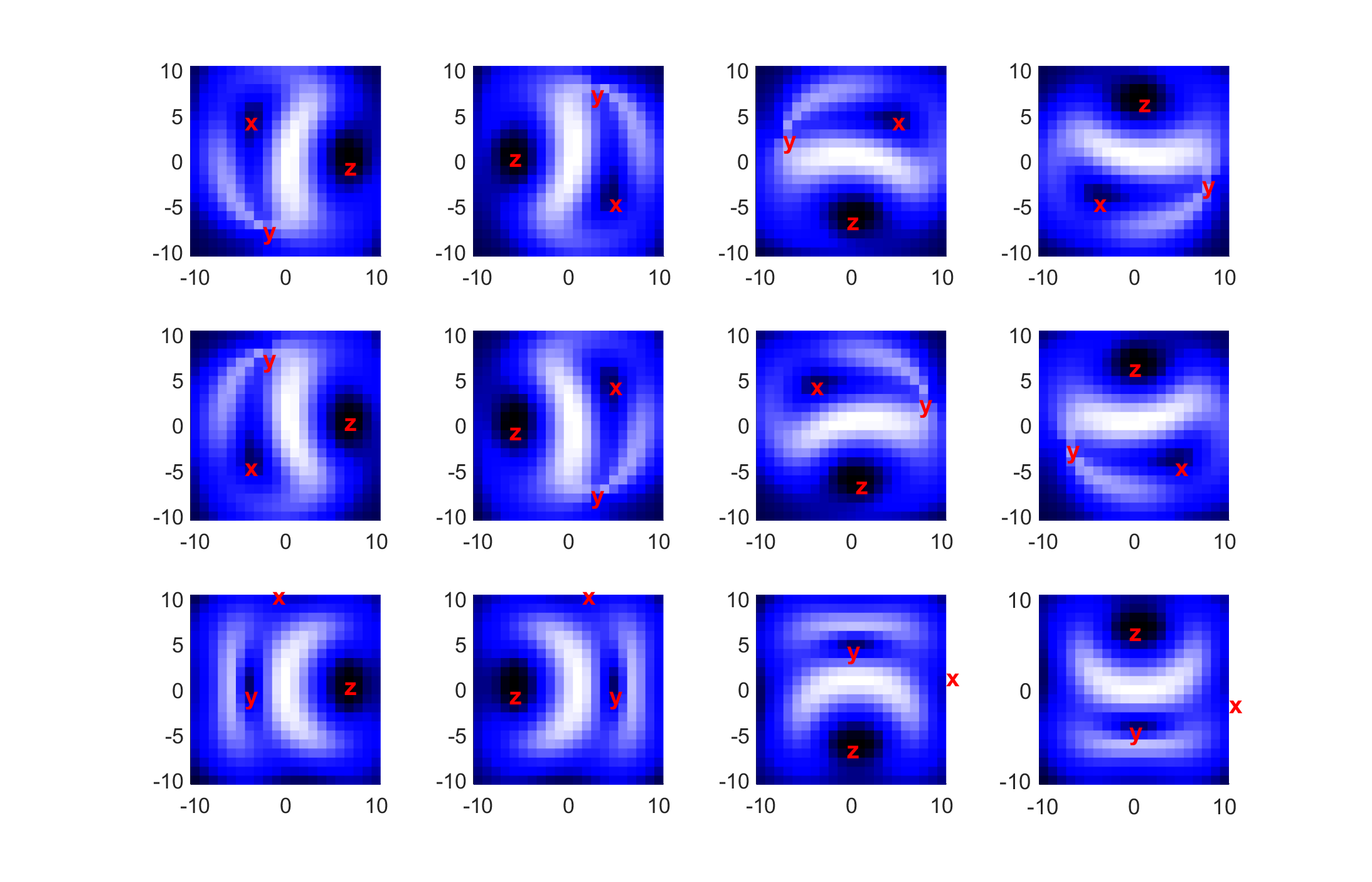}
          \includegraphics[width=\textwidth]{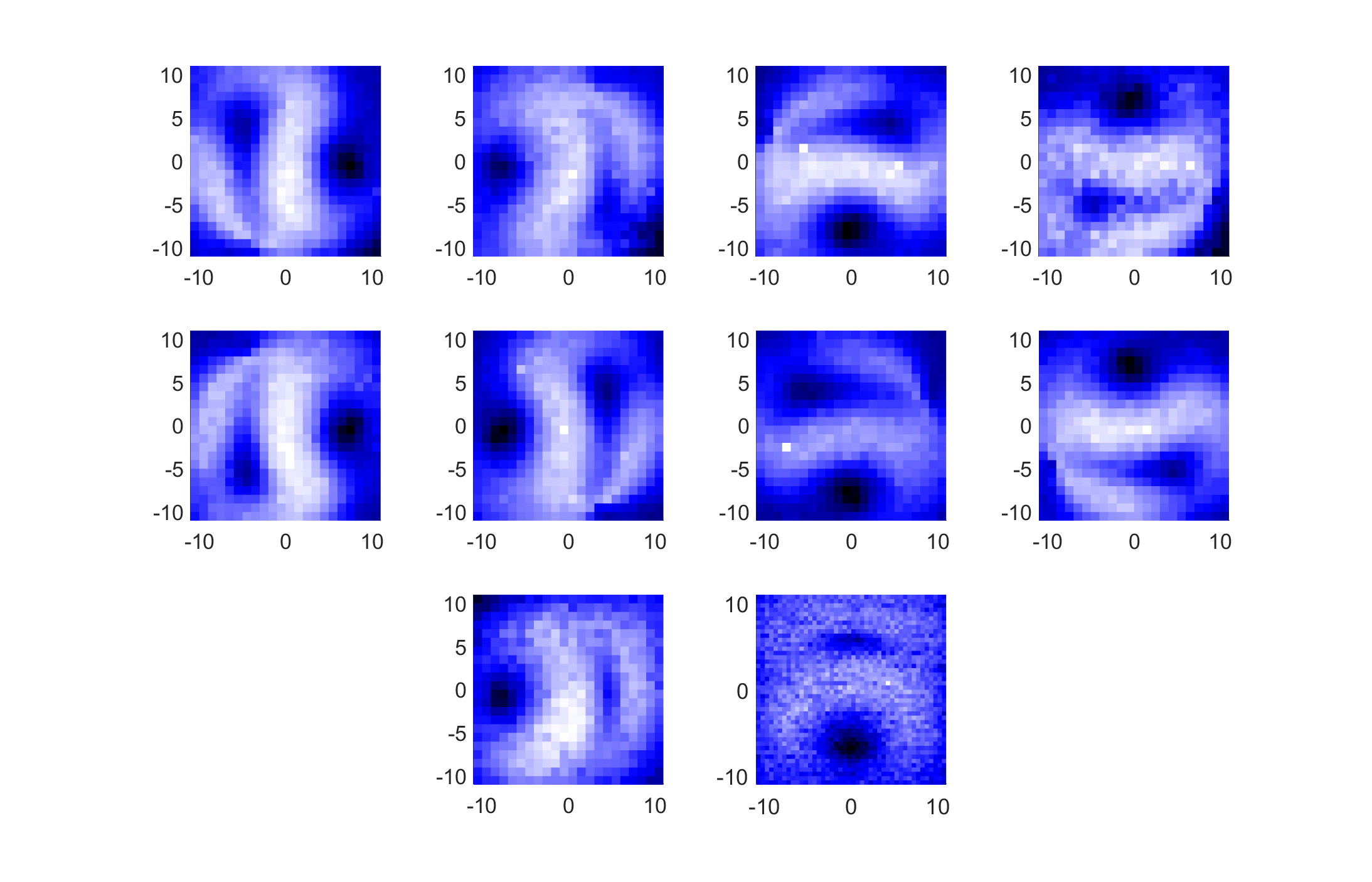}
     \caption{Simulated (top) and measured (below) magnetic maps for TR12. The axes display the magnet position in millimeters while the brightness is displayed by coloring. The two missing maps have not been observed so far.}
\end{figure*}
\newpage

\bibliography{LiteraturSup}